\documentclass[prd,aps,showpacs,nofootinbib,preprint,eqsecnum]{revtex4}
\usepackage{graphicx}
\usepackage[english]{babel}
\usepackage{amsmath}
\usepackage{amssymb}
\usepackage{amsfonts}
\usepackage{color,amsxtra}
\usepackage{epsf}
\usepackage{enumerate}
\usepackage{hhline}
\usepackage{array}
\usepackage{tabularx}
\usepackage{subfigure}
\usepackage{fancyhdr}
\usepackage{mathrsfs}
\newcommand{\be}{\begin{equation}}
\newcommand{\ee}{\end{equation}}
\newcommand{\bea}{\begin{eqnarray}}
\newcommand{\eea}{\end{eqnarray}}
\newcommand{\beaa}{\begin{eqnarray*}}
\newcommand{\eeaa}{\end{eqnarray*}}

\def\be{\begin{equation}}
\def\ee{\end{equation}}
\def\bea{\begin{eqnarray}}
\def\eea{\end{eqnarray}}

\begin{document}

\title{Proposal for the proper gravitational energy-momentum tensor}

\author{Katsutaro Shimizu}

\affiliation{Computer Science Division, The University of Aizu, Aizu-Wakamatsu 965-8580, Japan}


\begin{abstract}
We propose a gravitational energy-momentum tensor of the general relativity obtained using Noether theorem. It transforms as a tensor under general coordinate transformations. One of the two indices of the gravitational energy-momentum tensor labels a local Lorentz frame that satisfies the energy-momentum conservation law. The energies for a gravitational wave and a Friedmann--Lemaitre--Robertson--Walker universe are calculated as examples. 
\end{abstract}


\maketitle

\def\thesection{\Roman{section}}
\def\theequation{\Roman{section}.\arabic{equation}}

\section{Introduction}
Defining the gravitational energy-momentum tensor (GEMT) has been a long-standing problem in general relativity [1,2,3]. To derive an expression for the GEMT is fundamental and significant; its derivation would be very useful for other fields of general relativity. Several definitions of GEMT have been proposed. However, they are quite disparate and are not proper tensors [4,5,6,7,8,9].\par

In this paper we develop a proper GEMT using Noether theorem [10]. The important aspect introduced here is that in this theorem the general coordinate transformation must be written with tetrads. It induces the GEMT, which has an index for a local Lorentz frame that ensures the energy conservation law. \par

The organization of this paper is as follows. In Section II, we present a derivation of the GEMT via Noether theorem, in which the energy-momentum tensor depends on the volume terms and surface terms in the action. In general relativity, there are various physically different ways of decomposing the volume and surface terms. Mathematically, though, there are two different ways. Although the definition of the energy-momentum tensor of Noether theorem induces different formulae in the different decompositions, we can show that they are the same via identities.\par

The GEMT that is obtained by above method is not symmetric. Nevertheless, we acquire a symmetric GEMT using the conservation law for spin angular momentum tensor. When we consider a restricted region of space-time, we have to introduce Gibbons--Hawking term (GH term) [11] in the action. According to the GH term, we have to add extra terms in the GEMT. \par

In Section III, we evaluate the gravitational energies of a gravitational wave  and Friedmann--Lemaitre--Robertson--Walker(FLRW) universe [12,13,14,15]. The energy flow of a gravitational wave is well known although its expression is different in some of the literature [4,5,6,7,8,9]. We derive the energy flow and energy density of a gravitational wave.  In the FLRW universe, the sum of gravitational energy and energy of matter becomes zero everywhere.\par

Section IV is devoted to conclusions and discussion.\par

\section{Gravitational energy-momentum tensor in general relativity}

The action of general relativity is represented in a variety of ways as follows,
\begin{eqnarray}
I& =& \frac{1}{16\pi}\int d^4 x eR
\label{2.1} \\
&=&
\frac{1}{16\pi}\int d^4 x e (-\frac{1}{4}c_{\mu\nu\lambda}c^{\mu\nu\lambda}
-\frac{1}{2}c_{\mu\nu\lambda}c^{\lambda\nu\mu}
+c^{\mu\nu}_{\hspace{10pt}\mu} c^{\rho}_{\hspace{5pt}\nu\rho}+2\nabla_\mu c^{\nu\mu}_{\hspace{10pt}\nu} ) \label{TEGR}\\
&=&\frac{1}{16\pi}\int d^4 x e g^{\mu\nu}(\Gamma^\lambda_{\mu\rho}\Gamma^ \rho _{\nu\lambda}-\Gamma^\lambda_{\mu\nu}
\Gamma^\rho_{\lambda\rho})+\partial_\lambda e(g^{\mu\nu} \Gamma^{\lambda}_{\mu\nu}-g^{\lambda\nu}\Gamma^\sigma_{\sigma\nu}) \label{GR}\\
&=& \int d^4 x ( F+ \partial_\lambda D^\lambda ), \label{2.3}
\end{eqnarray}
where $e$ is the determinant of the tetrad $e^\mu_a$ and the gravitational constant $G$ has been set to be unity. Here, we have defined the tensor $c_{\mu\nu\lambda}\equiv e_{a\mu}(e^a_{\nu},_\lambda-e^a_{\lambda},_\nu)$ and $c^{\mu\nu\lambda}\equiv g^{\nu\rho}g^{\lambda\sigma}e_a^{\mu}(e^a_{\rho},_\sigma-e^a_{\sigma},_\rho)$, where the comma denotes partial derivative [16]. Latin indices signify local Lorentz coordinates, whereas Greek indices signify world coordinates. $F$ denotes volume terms and $\partial_\lambda D^\lambda$ denotes surface terms. Eq.~$(\ref{TEGR})$ is the teleparallel-equivalent of general relativity [17,18]. There are several ways to decompose the integrand into volume and surface terms. However, mathematically, every decomposition is equivalent to Eq.~$(\ref{TEGR})$ or Eq.~$(\ref{GR})$.\par

Under the following general coordinate transformation [19,20],
\begin{eqnarray}
 x^{'\mu}&=&x^{\mu}+\delta x^\mu \nonumber \\
         &=&x^{\mu}+e^{\mu}_a \xi^a,
\end{eqnarray}
the action $I$ is invariant. Here, $ \xi^a $ is an arbitrary function that is zero on the boundary. Accordingly, Noether theorem is satisfied as follows.
\begin{equation}
\partial_\mu(e^{\mu}_a \xi^a(F+\partial_\lambda D^\lambda))
+\delta^L F+\partial_\lambda \delta^L D^\lambda=0,\label{eq:2.5}
\end{equation}
where $\delta^L F $ and $ \partial_\lambda \delta^L D^\lambda $ are the Lie derivatives of $ F $ and $ \partial_\lambda D^\lambda $. It is used that the Lie derivative commutes with the usual derivative.\par

Specifically, the Lie derivatives of $ F $ and $ D^\lambda $ are
\begin{eqnarray}
\delta^L F&=&\frac{\partial F}{\partial e^a_\nu}\delta^L e^a_\nu+\frac{\partial F}{\partial e^a_\nu,_\mu}\delta^L e^a_\nu,_\mu, \label{eq:2.7} \\
\delta^L D^\lambda&=&\frac{\partial D^\lambda}{\partial e^a_\nu}\delta^L e^a_{\nu}+\frac{\partial D^\lambda}{\partial e^a_\nu,_\mu}\delta^L e^a_{\nu},_{\mu}.\label{eq:2.8}
\end{eqnarray}
and the Lie derivative of the tetrad is given by
\begin{eqnarray}
 \delta^L e^a_{\nu}&=&-e^a_{\sigma}\partial_\nu (e_b^{\sigma}\xi^b)
-e_b^{\sigma}\xi^b\partial_\sigma e^a_{\nu} \\
                  &=& -\xi^b(e_b^{\sigma}\partial_\sigma e^a_{\nu}
-e_b^{\sigma}\partial_\nu e^a_{\sigma})-\partial_\nu\xi^a \\
                   &=& -\xi^b c^a_{\hspace{5pt}\nu b}
-\partial_\nu\xi^a.\label{eq:2.10}
\end{eqnarray}
Substituting the Lie derivatives of Eqs.~$(\ref{eq:2.7}), (\ref{eq:2.8})$, and $ (\ref{eq:2.10})$ into the Noether theorem, Eq.~$(\ref{eq:2.5})$, we obtain
\begin{equation}
\partial_\mu(e t^{\mu}_a\xi^a-V_a^{\mu\nu}\partial_\nu\xi^a
-W_a^{\mu\nu\lambda}\partial_\lambda\partial_\nu\xi^a)
 +e T^\nu_a(\xi^b c^a_{\hspace{5pt}\nu b}+\partial_\nu\xi^a)=0, \label{noether}
\end{equation}
where $ T^\nu_a $ is the energy-momentum tensor of matter and we have used the equation of motion. Furthermore, $ t^{\mu}_a, V_a^{\mu\nu} $ and $ W_a^{\mu\nu\lambda} $ are defined as
\begin{eqnarray}
e t^{\mu}_a& \equiv& e^\mu_a(F+\partial_\nu D^\nu)
+\frac{\partial D^\mu}{\partial e_\nu^b}c^b_{\hspace{5pt}a\nu}
+\frac{\partial D^\mu}{\partial e^b_{\nu,\lambda}}\partial_\lambda c^b_{\hspace{5pt}a\nu}
+\frac{\partial F}{\partial e^b_{\nu,\mu}}c^b_{\hspace{5pt}a\nu}, \label{energy}\\
V_a^{\mu\nu} &\equiv& \frac{\partial F}{\partial e^a_{\nu ,\mu}}
+\frac{\partial D^\mu}{\partial e^a_\nu}
+\frac{\partial D^\mu}{\partial e^b_{\rho,\nu}}c^b_{\hspace{5pt}\rho a} ,\label{VVV} \\
W_a^{\mu\nu\lambda} &\equiv& \frac{\partial D^\mu}{\partial e^a_{\nu,\lambda}}
\label{2.14}.
\end{eqnarray}
As $\xi^a$ is the arbitrary function, Eq.~$(\ref{noether})$ reduces to four equations proportional to $\xi^a$, $ \xi^a_{,\nu}$, $\xi^a,_{\nu\mu}$, and $\xi^a,_{\mu\nu\lambda}$. They are
\begin{eqnarray}
\partial_\mu (e t^\mu_{\hspace{5pt}a})
+ e T^\nu_{\hspace{5pt}b}c^b_{\hspace{5pt}\nu a}=0,\label{det}\\
et^\mu_a-\partial_\nu V_a^{\nu\mu}+e T^\mu_a=0, \label{4.2}\\
(V_a^{\mu\nu}+\partial_\rho W_a^{\rho\mu\nu})\partial_\mu\partial_\nu\xi^a=0,\\
W_a^{\mu\nu\rho}\partial_\mu\partial_\nu\partial_\rho\xi^a=0.\label{W}
\end{eqnarray}
We easily find that $ V_a^{\mu\nu} $ is anti-symmetric with respect to $(\mu,\nu)$ in Eq.~$(\ref{TEGR})$. Therefore the derivative of Eq.~$(\ref{4.2})$ is written with
\begin{equation}
\partial_\mu e( t^{\mu}_{\hspace{5pt}a}+T^{\mu}_{\hspace{5pt}a})=0. \label{engyc}
\end{equation}
Moreover, from Eq.~$(\ref{GR}) $, $V_a^{\mu\nu}$ is not antisymmetric. Nevertheless $ (V_a^{\mu\nu}+\partial_\rho W_a^{\rho\mu\nu})$ is antisymmetric with respect to $(\mu,\nu)$. If we define $A^{\nu\mu}\equiv V_a^{\nu\mu}+\partial_\rho W_a^{\rho\nu\mu}$, Eq.~$(\ref{4.2})$ becomes
\begin{equation}
e(t^\mu_{\hspace{5pt}a}+T^\mu_{\hspace{5pt}a})
+\partial_\nu\partial_\rho W_a^{\rho\nu\mu}-\partial_\nu A^{\nu\mu} =0.
\end{equation}
Therefore the derivative of this equation also yields Eq.~$(\ref{engyc})$ in this case.\par

Because $ T^{\mu}_{\hspace{5pt}a} $ is the energy-momentum tensor for matter, it is natural to consider $ t^{\mu}_{\hspace{5pt}a} $ as the GEMT. Then Eq.~$(\ref{engyc})$ states the conservation law for the total energy-momentum tensor. This equation is covariant under general coordinate transformations because the energy-momentum tensor has only one world coordinate index. As the energy-momentum tensors are written usually with two world coordinate indices, the energy conservation law does not transform covariantly. General relativity has also another invariance, namely the global Lorentz transformation. However, this symmetry does not break Eq.~$(\ref{engyc})$.\par

We can derive the GEMT with Eqs.~$(\ref{4.2})$ and $(\ref{VVV})$, as well as the equation of motion. We obtain
\begin{eqnarray}
et^\mu_{\hspace{5pt}a}&=& \frac{1}{16\pi}(e^\mu_a e R
-2eR^\mu_{\hspace{5pt}a} +\partial_\lambda e(c_a^{\hspace{5pt}\lambda\mu}
-c^{\lambda\mu}_{\hspace{10pt}a}+c^{\mu\lambda}_{\hspace{10pt}a}))\\
&=&\frac{e}{16\pi}(e^\mu_a R-2R^\mu_a+\nabla_\lambda c_a^{\hspace{5pt}\lambda\mu}
+\nabla_\lambda c^{\mu\lambda}_{\hspace{10pt} a}
-\frac{1}{2}c_{a\lambda\sigma}c^{\mu\lambda\sigma}
+c_{\lambda\sigma a}c^{\lambda\sigma\mu}
+c_{\lambda\sigma a}c^{\sigma\lambda\mu} \nonumber \\
& & { }\qquad -\nabla_\lambda c^{\lambda\mu}_{\hspace{10pt} a}). \label{GEMT1}
\end{eqnarray}
The last term of the above equation is antisymmetric. We shall show that it is zero by the conservation law for the spin angular momentum tensor. For the other decomposition Eq.~$(\ref{GR})$, we obtain the same result, as well. It is well known that Eqs.~$(\ref{TEGR})$ and $(\ref{2.1})$ are equivalent. This result moreover ensures their equivalence .\par

The following identities are useful in a derivation.
\begin{eqnarray}
R^\mu_{\hspace{5pt}a}&=&\frac{1}{2}\nabla_\lambda(c^{\lambda\mu}_{\hspace{10pt} a}
-c^{\mu \lambda}_{\hspace{10pt}a}-c_a^{\hspace{5pt}\lambda\mu})
+\nabla_a c^{\lambda\mu}_{\hspace{10pt}\lambda}
-\frac{1}{2}c^{\sigma}_{\hspace{5pt}\lambda\sigma}(c^{\lambda\mu}_{\hspace{10pt}a}
-c^{\mu\lambda}_{\hspace{10pt}a}-c_a^{\hspace{5pt}\lambda\mu})\nonumber \\
&&{}+\frac{1}{2}c^{\mu\sigma\lambda}c_{\sigma\lambda a}
- \frac{1}{4}c^{\mu\sigma\lambda}c_{a\sigma\lambda}\\
&=& \frac{1}{2}\nabla_\lambda(c^{\mu\lambda}_{\hspace{10pt}a}
+ c_a^{\hspace{5pt}\lambda\mu})
+\frac{1}{2}(\nabla_a c^{\lambda\mu}_{\hspace{10pt}\lambda}
+\nabla^\mu c^\lambda_{\hspace{5pt}a\lambda}) +\frac{1}{2}(c^{\mu\nu}_{\hspace{10pt}a}
+c_a^{\hspace{5pt}\nu\mu}) c^\lambda_{\hspace{5pt}\nu\lambda} \nonumber \\
 & & {} +\frac{1}{4}(c^{\mu\sigma\lambda}c_{\sigma\lambda a}
+c_{a\sigma\lambda}c^{\sigma\lambda\mu})
-\frac{1}{4}c^{\mu\sigma\lambda}c_{a\sigma\lambda}.
\end{eqnarray}
and
\begin{equation}
\nabla_\nu c^\nu_{\hspace{5pt}a\mu}+\nabla_a c^\nu_{\hspace{5pt}\mu\nu}
+\nabla_\mu c^\nu_{\hspace{5pt}\nu a}
+c^\nu_{\hspace{5pt}\lambda\nu}c^\lambda_{\hspace{5pt}a\mu}
+\frac{1}{2}c_{\nu a\lambda}c_\mu^{\hspace{5pt}\lambda\nu}
+\frac{1}{2}c_{a\lambda\nu}c^{\nu\lambda}_{\hspace{10pt}\mu}=0.\label{id3}
\end{equation}
We comment on Eq.~$(\ref{det})$. From Eqs.~$(\ref{det})$ and $(\ref{engyc})$, we obtain
\begin{equation}
\partial_\mu (eT^\mu_{\hspace{5pt}a})=eT^\mu_{\hspace{5pt}b} c^b_{\hspace{5pt}\mu a}.
\end{equation}
Using the equation of motion, we can change $T^\mu_a$ by $ G^\mu_a$, the Einstein tensor. Then the above equation becomes
\begin{equation}
e(\partial_\mu G^\mu_a +\Gamma^\nu_{\nu\mu}G^\mu_a
-c^b_{\hspace{5pt}\nu a}G^\mu_b)=0. \label{bianchi}
\end{equation}
With Ricci's rotation coefficient given by
\begin{equation}
\Omega_{a b \mu}=-\frac{1}{2}(c_{a b \mu}+c_{b\mu a}-c_{\mu a b}),
\end{equation}
Eq.~$ (\ref{bianchi}) $ reads
\begin{equation}
\nabla_\mu G^\mu_{\hspace{5pt}a}=0.
\end{equation}
Here, the Bianchi identity can be derived.\par

Next, we consider the angular momentum conservation law of general relativity. To begin, we
consider a global coordinate rotation
\begin{equation}
x^{,\mu}=x^\mu-\omega^\mu_{\hspace{5pt}\nu}x^\nu,
\end{equation}
where $\omega_{\mu\nu}=-\omega_{\nu\mu}=const$. The Lie derivative of the tetrad is given by
\begin{equation}
\delta^L e^a_{\nu}
=\omega^\lambda_{\hspace{5pt}\sigma} x^\sigma\partial_\lambda e^a_\nu
+\omega _{\rho\sigma}x^\sigma e^a_\lambda\partial_\nu g^{\lambda\rho}
+e^a_\lambda \omega^\lambda_{\hspace{5pt}\nu}.
\end{equation}
Noether theorem becomes
\begin{equation}
\partial_\mu N^\mu=T_{\hspace{5pt}a}^\nu\delta^L e^a_{\hspace{5pt}\nu}.
\end{equation}
Here, $ N^\mu $ is Noether current, which is given by
\begin{equation}
N^\mu=-\omega^\mu_{\hspace{5pt}\nu} x^\nu eR
+(\frac{\partial F}{\partial e^a_{\nu,\mu}}
+\frac{\partial D^\mu}{\partial e^a_\nu})\delta^L e^a_{\nu}
+\frac{\partial D^\mu}{\partial e^a_{\nu,\lambda}}\partial_\lambda\delta^L e^a_{\nu}.
\label{angular}
\end{equation}
It is well known that the conservation laws of both the orbital and spin angular momentum tensors are derived from Noether　theorem. The conservation law for the spin angular momentum tensor is given by
\begin{equation}
\nabla_\mu c^{\mu a b}=0.
\end{equation}
This is the antisymmetric term of Eq.~$(\ref{GEMT1})$ when multiplied by a tetrad to change the local Lorentz coordinate into the world coordinate. Hence we obtain a symmetric GEMT, which reads
\begin{equation}
et^\mu_{\hspace{5pt}a}=\frac{e}{16\pi}(e^\mu_a R-2R^\mu_a+\nabla_\lambda c_a^{\hspace{5pt}\lambda\mu}+\nabla_\lambda c^{\mu\lambda}_{\hspace{10pt} a}-\frac{1}{2}c_{a\lambda\sigma}c^{\mu\lambda\sigma}+c_{\lambda\sigma a}c^{\lambda\sigma\mu}+c_{\lambda\sigma a}c^{\sigma\lambda\mu}). \label{GEMT4}
\end{equation}
\par

There is another symmetry in general relativity, this being the symmetry under global Lorentz transformations of the local Lorentz coordinate. Let us consider local Lorentz coordinate $ y^a(x)$. Under a global Lorentz transformation, the local Lorentz coordinate transforms as
\begin{equation}
y^{'a}(x)=y^a(x)+\omega^a_{\hspace{5pt}b}y^b(x),
\end{equation}
where $\omega_{a b}=-\omega_{b a}=const.$ The Lie derivative of the tetrad is
\begin{equation}
\delta^L e_\nu^a=-\omega^a_{\hspace{5pt}b}e^b_\nu.
\end{equation}
Noether theorem is now
\begin{equation}
\delta^LF+\delta^L\partial_\lambda D^\lambda=0.
\end{equation}
From the above equation we obtain a conservation law
\begin{equation}
\partial_\nu e(c^{\nu a b}+c^{a\nu b}-c^{b\nu a}
+e^{a\nu}c^{\lambda b}_{\hspace{10pt}\lambda}
-e^{b\nu}c^{\lambda a}_{\hspace{10pt}\lambda})=0,
\end{equation}
which is equivalent to Eq.~$(\ref{id3})$.\par

Next, we consider the GH term, which is introduced in the action when we consider a restricted region. This term is usually given by
\begin{equation}
        \frac{2\epsilon}{16\pi}\int d^3 x \sqrt {\gamma}(K-K_0),
\end{equation}
where $\gamma$ is the determinant of the 3-D space-time metric and $K$ is the trace of an extrinsic curvature. $K_0$ is the trace of the extrinsic curvature obtained from the flat space-time metric, and usually called the subtraction term. $\epsilon$ is $+1$ when the boundary is time-like and is $-1$ when the boundary is space-like. \par

The integration of $ K $ is expressed in terms of the tetrad as
\begin{eqnarray}
 \frac{2\epsilon}{16\pi}\int d^3 x \sqrt {\gamma} K&=&
-\frac{1}{8\pi}\int d^3x \sqrt{\gamma}N c^{j \zeta}_{\hspace{10pt}j} \nonumber\\
&=&-\frac{1}{8\pi}\int d^3x \sqrt{\gamma}\,n_\mu c^{j\mu}_{\hspace{10pt}j},
\end{eqnarray}
where $ n_\mu $ is an orthonormal vector of the boundary and $N$ is the lapse function. We assume the boundary to be a $\zeta=const.$ hypersurface. In this expression, it is not necessary to introduce $\epsilon$. To construct the GEMT from the GH term, we should treat the GH term as a volume term of a 3-D boundary and not a surface term of the 4-D space-time.\par

The effect of GH term in the GEMT is given by
\begin{equation}
t^i_\alpha(GH)
= -\frac{1}{8\pi}\frac{1}{\sqrt{|g^{\hspace{1pt}\zeta\zeta}|}}e^i_\alpha c^{j\zeta}_{\hspace{10pt}j},\label{GH1}
\end{equation}
and the effect of the subtraction term in GEMT is given by
\begin{equation}
t^i_\alpha(GH)_0
=-\frac{1}{8\pi}\frac{1}{\sqrt{|\bar{g}^{\hspace{1pt}\zeta\zeta}|}}e^i_\alpha\bar{c}^{j\zeta}_{\hspace{10pt}j}.\label{GH2}
\end{equation}
Here, $i$, $j$, and $ k$ are 3-D world coordinates and $\alpha$ and $\beta$ are 3-D local Lorentz coordinates on the boundary. In these equations, the $\zeta's $ are not summed. $\bar{c}^{j\zeta}_{\hspace{10pt}j}$ and $\bar{g}^{\hspace{1pt}\zeta\zeta} $ are  flat space time quantities.

\section{Gravitational energy}
In this section, we consider the GEMT in two examples, gravitational wave and FLRW universe.\par

As a first example, we consider the GEMT of the gravitational wave. The gravitational wave is usually considered as a weak plane wave in flat space-time. There are two modes in the gravitational wave, but for simplicity we consider only one mode moving in the z-direction. We take a transverse trace-less gauge. Therefore we can assume the metric has the form $g_{\mu\nu}=diag(-1,1+h_{xx},1+h_{yy},1)$. $ h_{xx}=Acos(kt-kz)=-h_{yy}$. As space-time has no boundary, we need not include the GH term in the GEMT. The gravitational wave is not an exact solution of the Einstein equation, but is a weak field approximation. Hence the Einstein tensor in the GEMT is written with $h_{xx}$ and $ h_{yy}$ as an approximation to second order. The $(t,t)$ component of GEMT is given by
\begin{equation}
et_{tt}=\frac{(kA)^2}{16\pi}(sin^2(kt-kz)-cos^2(kt-kz)+1/2).
\end{equation}
The energy density, namely, an average over a wave length of $ t_{tt}$ is
\begin{equation}
<et_{tt}>=\frac{(kA)^2}{32\pi}.
\end{equation}
The energy flow in the z-direction is written with $t_{tz}$. This aspect has been described in the literature [4,5,6,7,8,9], but differences among the results exist. The $et_{tz}$ is obtained as
\begin{equation}
et_{tz}=\frac{(kA)^2}{16\pi}cos^2(kt-kz).
\end{equation}
The average over a period is
\begin{equation}
<et_{t z}>=\frac{(kA)^2}{32\pi}.
\end{equation}
These are natural quantities.\par


As a second example, we consider the FLRW universe, which has the metric
\begin{equation}
ds^2=-dt^2+\frac{a^2(t)}{1-Kr^2} dr^2+(ar)^2 d\theta^2+(ar\sin\theta)^2 d\phi^2.
\end{equation}
The $ (tt) $ component of the GEMT for a FLRW universe is
\begin{equation}
t_{tt} = -\frac{3}{8\pi}(\frac{\dot{a}^2}{a^2}+\frac{K}{a^2}),
\end{equation}
where $ \dot{a} $ is a time derivative of a scale parameter; the equation of motion is
\begin{equation}
3(\frac{\dot{a}^2}{a^2}+\frac{K}{a^2})=8\pi\rho.
\end{equation}
Here, $\rho$ is the density of matter. The energy-momentum tensor for matter and $\rho$ are related by $T_{tt}=\rho$. Therefore
\begin{equation}
t_{tt}+T_{tt}=0
\end{equation}
holds. In a FLRW universe, the gravitational energy and the energy of matter cancel one another everywhere.

\section{Conclusions}
With the aid of Noether theorem, we have constructed a proper symmetric gravitational energy-momentum tensor of general relativity. The conservation law for the energy-momentum tensor holds. The gravitational energy-momentum tensor has a world coordinate index and a local Lorentz index. For this reason the energy-conservation law transforms covariantly under general coordinate transformations. The theories which have other local symmetries of space-time other than the general coordinate invariance, such as Poincare gravity [16], do not have a gravitational energy-momentum tensor as the other local symmetries break the energy conservation law. Researchers of teleparallel gravity have proposed a GEMT different from ours [17,18]. Their GEMT is not symmetric and the effect of the GH term is not included. Even if we start from the formula of teleparallel equivalent general relativity, Eq.~$( \ref{TEGR})$, our procedure leads to the same GEMT, Eqs.~$(\ref{GEMT4}), (\ref{GH1})$, and $(\ref{GH2})$.\par

In addition, we derived the gravitational energies of the gravitational wave and FLRW universe. The energy density and the energy flow of the gravitational wave are $(kA)^2/32\pi$.In the FLRW universe, gravitational energy and the energy of matter cancel each other out everywhere. \par

\section*{Acknowledgments}

The author sincerely thanks Professor Jiro Soda, Professor Ken-ichi Nakao, and Professor Akira Fujitsu for valuable discussions.

\end{document}